\documentstyle[11pt,newpasp,twoside,epsf]{article}
\def\Ha {H$\alpha$}

\def\etal {{\it\ et al}}
\def\eg {{\it e.g.}}
\def\ie {{\it i.e.}}

\def\edcomment#1{\iffalse\marginpar{\raggedright\sl#1\/}\else\relax\fi}
\reversemarginpar

\begin{document}
\title{Galaxies: The Third Dimension -- Conference Summary}
 \author{E. Emsellem}
\affil{Centre de Recherche Astronomique de Lyon, 69561 Saint-Genis Laval, France}
 \author{J. Bland-Hawthorn}
\affil{Anglo-Australian Observatory, PO Box 296, Epping, NSW 2121, Australia}

\section{Introduction}

``Galaxies: the Third Dimension'' continues a tradition which started in
Marseilles 1994 with the ``Tridimensional Optical Spectroscopic Methods
in Astrophysics'' meeting, followed by ``Imaging the Universe in Three
Dimensions'' at Walnut Creek in 1999.  We were both fortunate to attend 
all three meetings, and this review provides us with the opportunity
for retrospect. There have been developments both in instrumentation
and in the way astronomy is conducted, and these need to be seen against 
a broad canvas. As we emphasize below, one of the main reasons for
satisfaction is the trend towards talks focused on scientific results
rather than on `yet another 3D spectroscopic technology.'

\section{Cultural change}

For the benefit of future chroniclers, we note that the past
eight years has seen a number of cultural changes in how astronomy is
undertaken.  The forces of change include many things:

\begin{itemize}
\item the dominance of the web in the life of the modern researcher; 
the inexorable march of computer power, networking and software
development, with the prospect of a world-wide web-driven `virtual
observatory' looming on the horizon;
\item the researcher is rarely tied to one waveband or observing technique
-- the researcher has access to many instruments and data bases, and
is therefore spoilt for choice, and is less inclined to take on `raw data';
\item the burgeoning influence of active and adaptive optics on all large
telescopes; the trend towards 3D spectrographs over more traditional 2D
spectrographs; the steady rise in detector area and performance which
is the starting point of instrument design;
\item the commissioning of 8$-$10m telescopes and the development of
large expensive general purpose instruments catering to a broad 
constituency; 
\item traditional theorists now take far more interest in being
involved with observation and data -- theorist/experimentalist/observer
categories become more mingled to the benefit of more realistic
models and fleshier observational papers.
\item the changing culture imposed by funding for ground-based astronomy
through space agencies, and the stronger connection between space and
ground science; 
\item the weakening role of traditional observatories in producing 
science papers and the importance of research at universities; 
\item astrophysics has become the `new physics' due in part to the
cancellation of the Superconducting Super Collider in Texas, with the consequence that
many particle physicists have moved into astrophysics, an ever 
increasing number of faculty positions switched from physics into 
astronomy, and a rising fraction of astrophysics papers are moving
to the physics journals;  
\item the success of the microwave background explorers in determining 
in the world metric has established the importance of the universe as 
a laboratory -- this development has brought studies in galaxy
formation and evolution to the top of the agenda (Peebles 2000). 
\end{itemize}

The proceedings of the Marseilles, Walnut Creek and Cozumel meetings
show hallmarks of all of these factors. In ``Future Perspectives''
below, we try to foresee the future cultural trends.

\section{Meetings: Past and Present}

The Marseilles meeting (Comte \& Marcelin 1995) introduced the
astronomical community to the integral field spectrograph, first proposed
by G. Court\`{e}s. In just eight short years, we now find microlens arrays
in common use in almost all major observatories.
This is quite remarkable. The Conference Summary for
that meeting asked why long-slit spectrographs maintain their stronghold
in most observatories given that so much of the potential information
is thrown away. The instrumentation landscape is very different now.
The Marseilles meeting had a decidedly European flavour with a strong
focus on local heros C. Fabry and A. Perot.

Much the largest meeting of the three was held at Walnut Creek, California
under the auspices of Lawrence Livermore National Laboratories.
Not surprisingly, this meeting had a strong North American emphasis
and brought together most of the major instrument developers in the
western world.  This meeting saw the introduction of tunable filters,
the reemergence of image slicers, differential methods based on CCD
charge shuffling, energy-resolving 3D detectors, and a dizzying array
of enabling technologies for satellites and ground-based observatories.

The overall impression of the Cozumel meeting was the scientific maturity
with 3D spectrographs. Some groups have already achieved impressively
large survey samples with their devices. The meeting had a much stronger
focus on science than in previous meetings. At that time, astronomers
seemed overwhelmed with the problem of data analysis given that optical/IR
teams did not make the financial investment of the radio community in
software analysis tools. But fortunately this situation has changed with
observatories realizing the importance of a reduction pipeline. It is
well known that observatories (either in space or on the ground) which
deliver calibrated data to the user have the highest publication rates.

\section{Instrumentation}

``Galaxies: the Third Dimension'' brought together speakers from many
countries, with the largest contingent from Mexico and Spain. This is an
exciting time for Mexican-Spanish astronomy with the Grantecan 10.2m 
(GTC) due to see first light in late 2003. Many of the science talks,
which we review in the next section, showed that the Mexican-Spanish 
community was gearing up for Grantecan.
The first light instruments are the optical tunable filter/multi-slit
spectrograph Osiris, and the mid-infrared imager spectrograph 
Canaricam.  In the second phase, the GTC will have a near-infrared
imager/spectrograph called EMIR.

Osiris is modelled on the Taurus Tunable Filter (TTF) at the AAT,
having very similar parameters in all respects, although of course the
instrument will be mounted on a telescope with a sevenfold increase in
light-collecting power, and on a first rate observing site.  The entire
optical spectrum will be covered by a blue TF and a red TF.  Furthermore,
volume-phase holographic gratings will allow multi-slit spectroscopy
at low to mid resolutions. Another feature carried over from the TTF
is that all observing modes will be coupled with charge-shuffling which
allows for differential measurement (\eg\ perfect sky subtraction with
nod \& shuffle, perfect continuum subtraction with straddle shuffle,
etc.). Osiris will also allow for time series readout in either
spectroscopic or imaging modes.

We are at the dawn of an era for 3D spectrographs. All 8-10 meter
class telescopes will have such a device. The first ones
are already active or are being commissioned as we write (VMOS/VLT):
the Gemini suite with GMOS of course (Allington-Smith) but also the soon
to come GNIRS and NIFS (Miller), GIRAFFE/FLAMES on the VLT (Chemin) with
its 3 different modes, the double FP etalon on SALT (Reynolds). 
Most of these instruments are expected to open up scientific niches.

In an age of 3D spectrographs on the largest telescopes,
it is important not to forget that similar facilities on 4m class
telescopes are alive and kicking, and will continue to deliver
excellent and sometimes surprising science (e.g. LIFTS, OASIS, SAURON,
INTEGRAL, PMAS, various FP systems; examples provided by Wurtz, Arribas,
Bacon, Roth, Veilleux and many others). For the forseeable future, the
largest survey samples are going to arise from medium sized telescopes.
A particularly interesting
combination was demonstrated with GRIF (Clenet), which advantageously
combines the high spatial resolution attainable in the NIR with a 4m
aperture (see also Le Coarer), and the traditional scanning Fabry-Perot
technique. A related technique has been discussed in detail by Baldry
\& Bland-Hawthorn (2000).

A memorable exchange brought out the extremes of astronomy. Taylor
wowed the audience with the layout for the Californian Extremely Large
Telescope (CELT), and the proposed instrument suite -- `a billion here,
a billion there; soon this will start to add up to real money.' He
showed how HIRES could be rescoped for a 25m telescope.  Not to be
outdone, in a charming talk, Parry presented England's answer to the
CELT, the Cambridge Extremely Little Telescope, a 2~cm aperture
telescope with a fibre feed to CIRPASS, a spectrograph soon to see
first light on Gemini. The English CELT was observed to be `the largest
telescope in operation below sea level.'  Remarkably, Parry was able to
obtain the same sky flux per pixel as could be expected at Gemini, and
in the process demonstrate the excellent sky subtraction possible with
CIRPASS.

3D meetings are a good time to pause and consider the progress
made in the R\&D of true 3D detectors in the optical. With de Bruijne's
captivating talk, we heard of actual astrophysical applications of
superconducting tunnel junctions (4D detectors, adding a time
resolution of 5 $\mu$s), conducted at the WHT with a $6\times6$
elements system. The spectral resolution is very low ($R \sim 10$), but
already the reduction and analysis of the data flowing out of this
instrument is demanding.  A larger format device will require a
revolution in pipeline processing techniques. Talks by Charles and Javier
emphasized the importance of time resolved imaging and spectroscopy.

\section{Science with 3D spectrographs}

Multi-object studies have wide-ranging application, in particular,
star forming galaxies in clusters, around quasars, in QSO-identified
absorption line structures, and so on. The tunable filter studies
provide accurate photometry and object morphologies over the widest
possible field. The integral field studies provide broad spectral
coverage over a limited field. The compromise between spatial
and spectral field, or resolution, was nicely summarized by
Boulesteix: `it is now clear that astronomers do dig more freely
into the large bag of instruments available depending
on the scientific goal they wish to reach.'

Tunable filter, Fabry-Perot and integral field spectrographs together
then provide a powerful arsenal for going after a wide range of science
cases.  These can be broadly divided into two categories: detailed
studies of individual extended sources in absorption and emission;
multi-object spectroscopy of compact sources over a wide field. We
provide a brief overview of some of these topics before taking a closer
look at a few particularly promising areas.

There is a wide range of energetic processes which are expected
or known to produce detectable diffuse optical/IR line emission.
These include (in order of decreasing energy): colliding clusters;
cooling flows; gamma ray bursts, quasars (QSOs), radio galaxies
and ultraluminous IR galaxies (ULIGs); galaxy mergers; QSO/ULIG
jets and winds, AGNs and starbursts, hypernovae, supernovae,
compact x-ray sources, and so on. There exists a wide class of
exotic possibilities including flash photoionization events,
galaxy bow shocks and so forth. 

Many of these topics were covered by participants at the conference
(\eg\ van Breugel, Veilleux), with
a strong emphasis on objects with complex emission line distribution such
as in interacting systems: these are natural targets for spectrography
covering a two-dimensional field of view. Fabry-Perots are best suited 
to studies of a few spectroscopic lines over a large field, \eg\ H$\alpha$
to map the gas distribution and kinematics (Fuentes-Carrera, Plana,
Melo, Veilleux), although true IFS have also been extensively used to
probe merger remnants or galaxies with spatially extended bursts of star
formation (Arribas, Monreal-Ibero, Garc\'{i}a-Lorenzo).

Another trend we are seeing is the realization of the importance of
detailed astrophysical studies of nearby sources. This is not as obvious
a statement as it sounds. Until recently, many astronomers would have
preferred to keep the objects as simple as possible by viewing them
at the highest possible redshift. But important physical processes are
now evident in pixel-pixel comparisons over several spectral diagnostics.
A superb example is the bipolar wind in M82. The individual emission line
maps are extremely complex and most observers would dismiss this information
as weather, \ie\ unimportant detail. However, Shopbell \& Bland-Hawthorn
(1998) show that the ratio of, say, [NII]/\Ha\ reveals a very well 
organized bipolar fans where the UV radiation is escaping from the 
galaxy core.

3D spectrographs provide detailed information on more than one physical
measurement at a time and are thus efficient at delivering the key
diagnostics over a spatial region freed from simple geometrical
assumptions.  It is then critical to understand how the spatial
averaging on our target affects our observational data. Talks by
Points, Rosado, Silich are clearly telling us that this is a complex
issue, specially when dealing with the ISM and star forming regions.
The multi-wavelength approach is certainly a key here, but without the
two-dimensional spatial coverage, such spectroscopy would look like a
list of street names without a city map.

`Feedback' is probably one of the most overused words in this context,
probably because it seems to include a subset of physical processes
depending on the author and the field.  Winds and radiation from celestial sources
provide the mechanisms for feedback, but wind energetics are notoriously
difficult to measure directly.  A courageous attempt to map the impact
of winds due to star formation was presented by Hidalgo-G\'amez, who
scanned the dwarf but starbursting IC 10 with long-slits. But this was
not counting on the power of their galactic (super)versions,  beautifully
illustrated by Veilleux's review, who convincingly argued for the need
of 3D spectrographs but with high spatial resolution.

We have also seen quite a number of detailed studies on bright nearby
galaxies, again mainly dealing with the gas distribution and kinematics
(Cair\'os, Puerari, Rela\~no Pastor, Valdez-Gutierrez). As we move
towards the central cores of these galaxies, more irregularities,
asymmetries and overlapping physical states were revealed
(Garc\'{i}a-Lorenzo, Mediavilla, Moiseev).

The success of 3D spectroscopy in obtaining absorption line maps of
galaxies shows us how to open a large window on the study of stellar
populations and kinematics in extended and sometimes remote objects.
As shown by many contributers (Sil'chenko, Moiseev, Bacon, Emsellem,
Gonz\`alez, Bureau, Cretton, Wernli, Falc\'on Barroso), the real
difficulty is to reconstruct the link between the stellar kinematics and
underlying populations.  State of the art dynamical models are required
(Cretton, Pichardo, Martinez), but we still need a `coloured' version of
these models if we really wish to build a good scenario for the formation
and evolution of these objects.

As mentioned above, survey science with 3D spectrographs is now of
increasing importance. This is typified by the GHASP (FP, Amram),
WHAM (FP, Reynolds) and Sauron (IFS, Bacon) surveys. 

The Sauron team (Bacon, Bureau, Cappellari, Copin, Cretton, Emsellem,
Falc\'on Barroso, Wernli) presented integral field results for a large
sample of elliptical, lenticular and spiral galaxies divided between
clusters and the field. Fully one third of these show evidence of
nuclear sub-structure. Some show evidence for triaxility, including
nuclear bars as judged from Athanassoula's bar simulations.  The group
emphasizes caution in the analysis and derivation of black hole masses
from long-slit measurements.

Some of the most impressive conclusions were those derived from the
age and abundance information coupled to new stellar synthesis models
(Vazdekis).  The new models, which have a
fourfold increase in spectroscopic resolution compared to the Lick
system, show that the isochrone or isochemical grid lines overlaid on a
plot of two Lick indices are more orthogonal than the Worthey models.
Thus, galaxies like NGC~4365 that exhibit no age gradient in the
Vazdekis models (Davies\etal\ 2001) appear to
show an age spread in the Worthey models. Interestingly, NGC~4150
exhibits an abundance spread with constant age.

It will probably take time before astronomers can fully exploit the
potential of such data sets. But one thing is clear from the Sauron
talks: axisymmetry in early-type galaxies is the
exception rather than the rule -- these are not quite the simple 
systems the community was led to believe.

\section{Software}

A traditional argument used {\em against} 3D spectroscopy was
simply that `the data are too complex, too difficult to model, and
impossible to calibrate.'  Most institutes carrying out the
construction of 3D spectrographs were well aware of the importance of
software and more specifically of semi-automated reduction pipelines. The
Cozumel meeting was an opportunity to witness the results of these
efforts. Dedicated IRAF tasks have been developed for the GMOS IFU on Gemini
and presented by Miller. Garcia-Lorenzo focused on IDA, an IDL based soft
tool to view and analyse INTEGRAL/WHT data, and the OASIS \& SAURON teams
showing some examples of reduction and analysis via the corresponding
(C-based) XOasis and XSauron.  We have also seen various applications
and extensions such as 3D deconvolution or 2D binning (Cappellari).

The GHASP survey (Amram) is a good illustration of how Fabry-Perot
H$\alpha$ velocity fields can now be produced in reasonably large
numbers. We are thus starting to really explore the techniques which will
truly exploit the full three-dimensional data at our disposal.  But more
importantly, astronomers are getting used to the concept of datacubes. In
this context, Brinks is right to remind us that radio astronomers have
been tackling 3D data for a few decades already.  Some new developments
are and will certainly be inspired from existing tools, like the analysis
of two-dimensional maps with the newly baptised ``kinemetry'' (Copin).
We however feel that the current trend really goes one step beyond,
since it appears to be providing new algorithms and tools for a much
wider net of applications than before.

Although the existing software packages make use of a zoo of different
languages and platforms, the community is actively engaged in establishing
a solid and common basis for future developments.  The advent of a
European format for spectroscopic datacubes perfectly illustrates this
point. The Euro3D network is just starting, so we are looking forward
to harvest the fruits of these collaborations at the next 3D meeting.

\section{Future Perspectives}

It continues to surprise us that so little attention has been paid to the
`diffuse light' universe. We know that there must be a lot of activity
at low surface brightness which is not seen in contemporary images. A
customized machine would be demanding to build. An off-axis telescope
would bypass the problem of the diffraction pattern. The mirror needs to
be kept clean of dust and the optics well specified and AR coated. This
leaves the not insignificant problem of atmospheric scattering. A 5th
mag star can be detected a degree or more away. This of course can be
bypassed from a satellite platform.

Surveys drive astronomy at any wavelength. We have seen a gradual trend
towards field widening of telescopes and instruments. This raises the
problem of how to sustain diffraction limited imaging or spectroscopy
over a wide field. Gemini has invested much effort in multi-conjugate
adaptive optics, and the second generation of VLT instruments will
certainly address this too. An alternative approach might be deployable
adaptive optics correctors connected to deployable IFUs or microIFUs.

The strategic initiative of the Next Generation Space Telescope has
provided the impetus for numerous enabling technologies relating to 
multi-object spectroscopy. Complex fibre positioning systems would
be far too risky for a remote space observatory. 
Some of the ideas under consideration are transmissive spectrographs which use
2000$\times$2000 micro-shutters and would allow for fully programmable
electronic object selection keeping the number of moving parts at
a low level. 

We predict that there will be a complete blurring of the traditional
wavelength boundaries. It will however take time before we can digest
the wealth of information contained within the flowing stream of 3D
data. This adaptation will certainly represents a higher number of
generations for astronomers than for computers.  Another prediction,
or rather a strong wish, is to reach the state of fully automated data
analysis more or less directly linked to virtual observatories. A first
step is being taken to obtain a standard (and optimized) format at the
output of the 3D spectrographs. 

The rising cost of future telescopes and instrumentation carries the
risk that astronomy will price itself out of existence.  We may see the
return of specialist `niche' instruments which target a specific science
question.  The Dazle instrument (Cambridge/AAO) is an example of this may
change. The instrument is entirely optimized for high redshift Ly$\alpha$
imaging, and shares an expensive camera with the CIRPASS spectrograph. The
camera is a modular unit with handles and can be ferried back and forth
between the instruments with little fuss.

Both software and instrumentation developers will need to be mindful
of spiralling costs. Large survey machines like the Sloan Telescope
were almost undone by the enormous expenditure involved in pipeline
reduction. Huge massively multiplexed spectrographs offer many more
features than are likely to find effective use in the ten to fifteen
year lifetime of the instrument. Now more than ever, if in doubt,
astronomers really must focus on a particular science niche, where 
all other scientific considerations are secondary, and remain that
way until the instrument sees first light. Even if the 
instrument is intended for the general community, there must be
unique terrain for the instrument to explore if it is to avoid the
fate of being `yet another 3D spectroscopic technology.'

\acknowledgements We would like to thank the conference organizers for arranging an
excellent meeting at such a resplendent venue in Cozumel just a short
distance from the Yucatan peninsula. Quite apart from hearing about the
latest developments in astronomical 3D instruments and observations, this
afforded the opportunity for some wonderful diving. We would both like
to thank Drs. Bureau, Cretton, Miller, Rela\~{n}o-Pastor and Veilleux
for the memorable shared experiences among the gardens and canyons 80
feet below. And then there were the endless `happy hours' at the pool
or on the beach, dance parties till sunrise, all in the interests of
international relations.

\end{document}